\documentstyle[aps]{revtex}
\input{epsf}

\newcommand{\ma}[4]
  {\left( \begin{array}{cc}
          #1 & #2 \\
          #3 & #4
   \end{array} \right)}

\newcommand{\ud}{\textrm{d}}

\newcommand{\GA}{\Gamma}

\begin{document}

\title{\bf{Random matrix theory for systems with an approximate symmetry \\ 
           and widths of acoustic resonances in thin chaotic plates}}

\author{A. Andersen$^1$, C. Ellegaard$^2$, A. D. Jackson$^2$, 
        and K. Schaadt$^2$ \\ 
       {\it $^1$The Technical University of Denmark, Department of Physics, 
                DK-2800 Lyngby, Denmark, and Ris\o{} \\ National Laboratory,
                Optics and Fluid Dynamics Department, DK-4000 Roskilde, 
                Denmark. \\
            $^2$The Niels Bohr Institute, Blegdamsvej 17, DK-2100 
                Copenhagen \O, Denmark} \\
(\today) \\
\begin{abstract}
\begin{quotation}
We discuss a random matrix model of systems with an approximate symmetry and 
present the spectral fluc\-tua\-tion statistics and eigenvector 
characteristics for the model. An acoustic resonator like, e.g., an aluminium 
plate may have an approximate symmetry. We have measured the frequency 
spectrum and the widths for acoustic resonances in thin aluminium plates, cut 
in the shape
of the so-called three-leaf clover. Due to the mirror symmetry through the 
middle plane of the plate, each resonance of the plate belongs to one of two
mode classes and we show how to separate the modes into these 
two classes using their measured widths. We compare the spectral statistics of 
each mode class with 
results for the Gaussian orthogonal ensemble. 
By cutting a slit of increasing depth on one face of the plate, we
gradually break the mirror symmetry and study the transition that takes place
as the two classes are mixed. Presenting 
the spectral fluctuation statistics and the distribution of widths for the 
resonances, we find that this transition is well described by the random 
matrix model.
\vspace{0.4cm}
\end{quotation}
\end{abstract} }

\maketitle

\section{Introduction}

Random matrix theory has been used with success in a variety of 
physical systems for the description of certain generic features of 
spectral correlators which are determined by the underlying symmetries 
of the Hamiltonian \cite{review}. In Sec.\ II of this paper we 
discuss a random matrix model of sy\-stems with an approximate symmetry. 
A problem like this is found, e.g., in nuclear physics where isospin 
symmetry, characteristic of the strong interactions, is only approxi\-mate 
due to Coulomb effects \cite{mitchell}. Isospin mixing was analysed by 
Guhr and Weidenm\" uller in 1990 using a random matrix approach \cite{guhr}. 
They used a random matrix  model to describe experimental data and to 
estimate the ave\-ra\-ge symmetry-breaking ma\-trix element, i.e., the 
ave\-rage Coulomb matrix element. The random matrix model discussed here 
differs from the one considered in \cite{guhr}, and we comment on this 
dif\-ference. In addition to the spectral fluctuation statistics for the model 
we consider a measure of the asymmetry of the eigenvectors and describe it 
using simple analy\-ti\-cal arguments. 

In Sec.\ \ref{acoustic.sec} we present two experi\-mental studies of 
acou\-stic resonances in thin aluminium plates. The plates have the shape of 
the so-called three-leaf clover, see 
Sec.\ \ref{acou_sys.sec}. Frequency spectra of acoustic resonators were first 
compared with random matrix results by Weaver in 1989 \cite{weaver}. Further 
experimental studies of the fluctuation properties of acoustic resonance 
spectra in blocks of aluminium and quartz were made by Ellegaard and 
coworkers \cite{acoustic1:1995,acoustic2:1996}. In Ref. \cite{leitner} the 
level spacing distribution measured in \cite{acoustic2:1996} was compared 
with the random matrix model of \cite{guhr}. In this paper we 
focus on acoustic plates which in many respects are simpler 
than the three-dimensional reso\-nators mentioned before.
Acoustic resonances in plates were investigated 
by Bertelsen et al. \cite{Hugues:2000}. We present a short review of 
the theory of acoustic waves in thin isotropic plates and discuss the 
characteri\-stics of the different types of resonances. We find 
expe\-ri\-mentally 
that modes can be separated into two dif\-ferent classes which each have a 
cha\-rac\-teristic dependence of their widths on the damping by the air
surrounding the plate. One class of modes has widths which are almost
independent of the air-pressure, and the other class has widths with a strong
dependence on the air-pressure. We argue that these modes are
in-plane and flexural modes, respectively. In the first experiment, we measure
the spectral fluctuation statistics for both mode types individually and
compare with well-known results for the Gaussian orthogonal ensemble (GOE). 
Then, in a second experiment, we mix the two mode classes by gradually 
cutting a slit on one face of the plate. We thus observe the transition from
two separate classes of modes to one class of modes. This transition is 
studied by comparing the data to the random matrix model for sy\-stems with 
an approximate symmetry for both the spectral fluctuation statistics and the 
width distribution. The latter is described using eigenvector 
information from the model.

\section{Random Matrices and Approximate Symmetries}
\label{theory.sec}

\subsection{The random matrix model} 

Let $H$ be a random real symmetric $N \times N$ matrix with the following
block-structure:

\vspace{0.4cm}

\begin{equation}
\label{goediagonal}
H \ = \ \ma{D_{A}+A}{0}{0}{D_{B}+B} \ + \ g \ \ma{0}{C}{C^{T}}{0} \ ,
\end{equation}

\vspace{0.4cm}

\noindent
where $D_{A}$ and $A$ are random $N_1 \times N_1$ matrices, and $D_{B}$ and 
$B$ are random $N_2 \times N_2$ matrices. The random matrix $C$ is 
$N_1 \times N_2$, and the coupling strength, $g$, is a real parameter. Note
that $N \equiv N_{1}+N_{2}$. The elements of the diagonal matrices $D_{A}$ 
and $D_{B}$ are drawn uniformly on the interval $[-0.5,0.5]$ and ordered in 
increasing order for each block. This choice of probability distribution leads 
to level spectra for $D_A$ and $D_B$ which, except for small end-point 
corrections, are described by the Poisson statistics appropriate for a 
sequence of uncorrelated energy levels. The elements of $A$, $B$, and $C$ are 
Gaussian distributed
with zero mean. The variance, $\sigma^{2}$, of the distribution of the 
diagonal elements of $A$ and $B$ scales as $1/N^{2}$. The variance of the 
distribution of the off-diagonal elements of the matrices $A$ and $B$ and 
the elements of $C$ is set to half the value of $\sigma^{2}$.

The diagonal contributions $D_A$ and $D_B$ in $H$ are intended to 
mimic the effects of the kinetic energy operator, and the Gaussian 
distributed elements of $A$ and $B$ simulate ``interactions'' due to 
boundary conditions. Since 
the ele\-ments of $D_{A}$ and $D_{B}$ are ``sufficiently'' small compared with 
the Gaussian distributed elements, the short-range spectral fluctuation 
statistics are identical to the statistics obtained for two superimposed GOE 
spectra (2 GOE) when $g=0$ and to GOE statistics when $g=1$. (See Sec.\ 
\ref{spec_fluc_stat} for a more detailed discussion of the spectral
fluctuation statistics.) The ave\-rage di\-stan\-ce between neighbouring 
levels scales as $1/N$ because of the presence of the diagonal matrix 
elements. With a finite value of $g$ both the root mean square (RMS) 
symmetry-preserving matrix element and the RMS symmetry-breaking matrix 
element also scale like $1/N$, and the transition from 2 GOE statistics to GOE 
statistics takes place as a function of $g$ independent of the value of $N$.

This is not the case if the kinetic energy terms are not present as in a 
random matrix model, like the one used in Refs. \cite{guhr,leitner}, with 
two GOE-like diagonal blocks coupled by Gaussian-distributed matrix elements. 
For such a model the ratio between the RMS symmetry-breaking matrix element 
and the average di\-stance between neigh\-bouring levels for the unperturbed 
problem scales like $g \sqrt{N}$. The transition from 2 GOE to GOE spectral 
fluctuation statistics takes place as a function of this ratio. If $g$ is 
independent of $N$, it follows that the ratio scales like $\sqrt{N}$, and in 
particular that it goes to infinity in the large-$N$ limit for any finite 
value of $g$. To observe a smooth transition from 2 GOE statistics to GOE 
statistics independent of $N$, it is thus necessary that the ratio between 
the RMS  symmetry-breaking  matrix ele\-ment and the RMS symmetry-preserving 
matrix ele\-ment scales like $1/\sqrt{N}$ in a random matrix model without 
kinetic energy terms.

\subsection{Spectral fluctuation statistics}
\label{spec_fluc_stat}

\begin{figure}
\centerline{\epsfxsize=8.5cm
            \epsffile{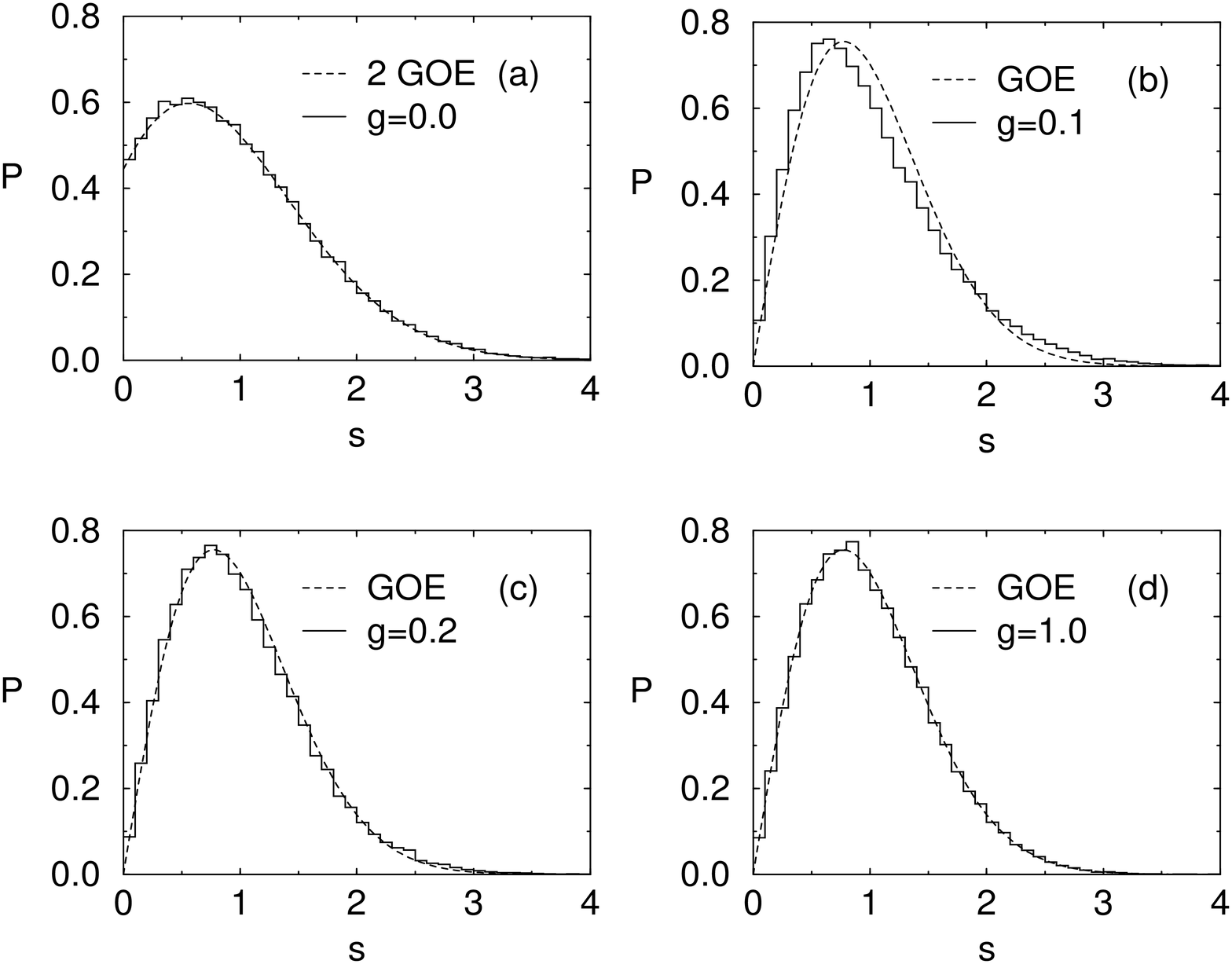}
           }
        \vspace{0.3cm}
        \caption{The level spacing distribution for different values of $g$. 
                 The two diagonal blocks are uncoupled when $g=0$, and when 
                 $g=1$ all the basis states are coupled.}
        \label{levelspacing.fig}
\end{figure}

\begin{figure}
  \epsfxsize=8.5cm
  \epsfysize=6.4cm
  \centerline{ {\epsffile{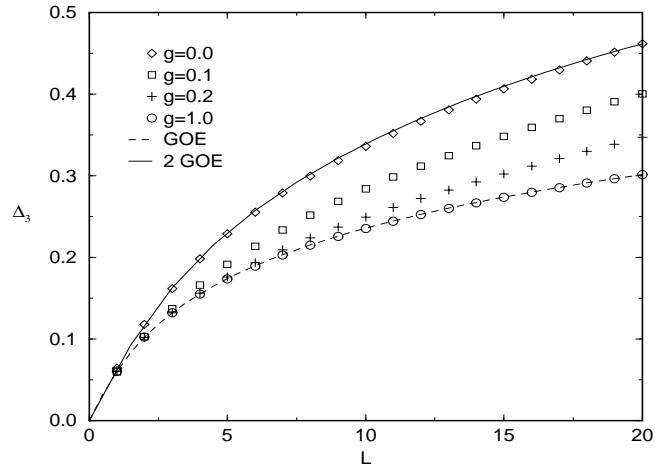} } }
  \caption{The $\Delta_{3}$-statistics for different values of $g$. The 
           numerical simulations for $g=0$ and $g=1$ agree with the
           2 GOE and the GOE result, respectively. }
  \label{delta3.fig}
\end{figure}

To describe the short-range spectral fluctuation statistics we consider the 
standard level spacing distribution, and as a measure of the 
long-range spectral fluctuation statistics we choose to look at the 
$\Delta_{3}$-statistic \cite{mehta}. Numerical calculations of the level 
spacing distribution and the $\Delta_{3}$-statistic for $N_{2}=2N_{1}=200$ 
with $\sigma^2 = 16/N^2$ are shown in Figs.\ \ref{levelspacing.fig} and 
\ref{delta3.fig} together with 
the exact results for the GOE and two superimposed GOE spectra with 
fractional densities 1/3 and 2/3, respectively. The ensembles in the
simulations consisted of 500 matrices, and 150 eigenvalues from the ``middle'' 
of the spectrum for each matrix
were con\-si\-dered. Figures 1(a) and 1(d) show that the level spacing 
distribution for the random matrix model is identical to the 2 GOE result when 
$g=0$ and to the GOE result when $g=1$. Similarly Fig.\ \ref{delta3.fig} 
shows that this is also the case for $\Delta_{3}(L)$ for $L \leq 20$. It 
is clear from Fig.\ \ref{levelspacing.fig} that the level spacing distribution 
looks very much like the level spacing distribution for the GOE even when 
$g=0.1$. It is well known that $\Delta_{3}(L)$ for a 
model with diagonal terms like $D_{A}$ and $D_{B}$ deviates from the 
corre\-spon\-ding GOE result for large values of $L$ \cite{guhr:1989}. 
The value of $L$ where this transition to more Poisson-like behaviour sets 
in is referred to as the Thouless energy. For $g=1$ and $\sigma^2=16/N^2$ we 
find a Thouless energy of about 35. A different choice of the variance, 
$\sigma^{2}$, leads to a picture for the spectral fluctuation statistics 
similar to the one shown on Figs.\ \ref{levelspacing.fig} and 
\ref{delta3.fig} as long as $L$ is less than the Thouless energy.

\subsection{Eigenvector information}

As a measure of the asymmetry of the eigenvectors of $H$ we define a quantity, 
$a$, which we denote the asymmetry number. Consider a $N=N_1+N_2$ dimensional 
vector $(v_{1}, v_{2}, ..., v_{N})$ of unit length, and let $a$ be defined by

\begin{equation}
a \ \equiv \ \sum_{i=1}^{N_1}v_{i}^{2} - \sum_{i=N_1+1}^{N}v_{i}^{2} \ .
\end{equation}

\begin{figure}
\centerline{\epsfxsize=8.5cm
            \epsffile{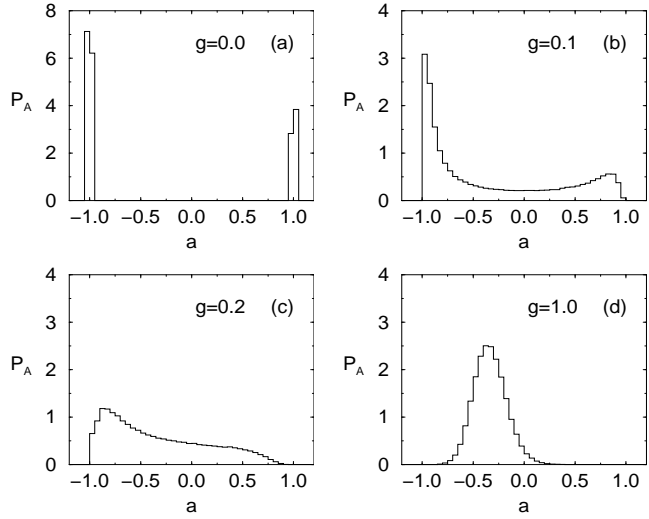}
           }
        \vspace{0.3cm}
        \caption{The asymmetry distribution $P_{A}$ for the
                 eigenvectors. When $g=0$ each eigenvector belongs to one
                 of the two parts of the vector space. As $g$ increases the
                 eigenvectors get components in both of the two parts of 
                 the vector space.}
        \label{asymmetry.fig}
\end{figure}

\noindent
For two decoupled systems described by the subspaces spanned by the first 
$N_1$ and the last $N_2$ basis vectors, respectively, the distribution 
$P_{A}(a)$ has a $\delta$-function peak at $a = -1$ and one at $a = 1$. For 
the GOE it has a single peak at $a = (N_{1}-N_{2})/N$. These features are 
obvious in Fig.\ \ref{asymmetry.fig}, which shows $P_{A}$ calculated 
numerically for the ensembles considered in Sec.\ \ref{spec_fluc_stat}. 
Notice that the smaller of the two peaks present when $g=0$ has almost 
vanished when $g=0.1$, whereas the strength of the largest peak is reduced to 
half its original value. 

Imagine that two uncoupled classes of resonances have width distributions
$P_{F}$ and $P_{I}$, respectively. The width distribution of all the 
resonances, $P(\GA )$, is the sum of $P_{F}$ and $P_{I}$ when the two classes
are uncoupled. The width distribution, $P(\GA )$, changes if the two classes
are coupled, and in our random matrix approach we model $P(\GA )$ using the 
asymmetry distribution:

\begin{eqnarray}
\label{width.eqn}
  P(\GA)     &=& \int_{0}^{1} \ud x
                 \int_{-\infty}^{\infty} \ud \GA_{F}
                 \int_{-\infty}^{\infty} \ud \GA_{I} \\
    & & \hspace{-1.2cm} \times
                 \ \delta(\GA- x \GA_{F}- [1-x] \GA_{I})
                 \ 2 P_{A}(1-2x) P_{F}(\GA_{F}) P_{I}(\GA_{I})
                 \nonumber \\
  = & & \hspace{-0.6cm}  
                  \int_{0}^{1} \ud x \frac{2 P_{A}(1-2x)}{1-x}
                  \int_{-\infty}^{\infty} \ud \GA_{F}
                  P_{F}(\GA_{F})P_{I}\left(\frac{\GA-x\GA_{F}}{1-x}\right) 
                  \nonumber \ .
\end{eqnarray}

\noindent
Notice that the integral reduces to the weighted sum of $P_{F}$ and $P_{I}$ 
if $P_A$ is the sum of two $\delta$-functions as in the case $g=0$ shown 
in Fig.\ \ref{asymmetry.fig}(a), and that $P$ is expressed in terms of $P_A$ 
if $P_{F}$ and $P_{I}$ are $\delta$-functions.

\begin{figure}
\centerline{\epsfxsize=8.5cm
            \epsffile{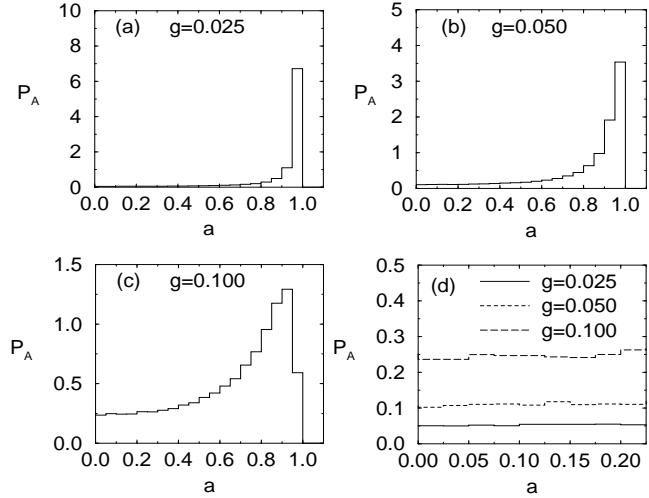}
           }
        \vspace{0.3cm}
        \caption{The asymmetry distribution $P_{A}(a)$ obtained numerically
                 for random matrix ensembles with $N_{1}=N_{2}$. The value  
                 $P_{A}(0)$ grows linearly with $g$, see the blow-up on (d), 
                 as 
                 predicted in the simple analytical model of the eigenvector 
                 mixing, see Eq. (\ref{pa_dist}).}
        \label{asymmetry_1:1.fig}
\end{figure}

We now consider the case $N_{1}=N_{2}$ and describe the characteristic
properties of the asymmetry distribution using a simple analytical model and
arguments from perturbation theory. Figure \ref{asymmetry_1:1.fig} shows
numerical calculations of the distribution of the asymmetry numbers for 
$N_{1}=N_{2}=150$ for three values of $g$. We considered ensembles of 500
random matrices, and, as in the numerical simulations described in Sec.\ 
\ref{spec_fluc_stat}, we focused on the 150 eigenvalues in the ``middle'' of 
the spectrum. The distribution $P_{A}(a)$ increases linearly as a 
function of $g$ close to $a=0$ as shown in Fig.\ \ref{asymmetry_1:1.fig}(d).  

The small fraction of the eigenvectors for which the asymmetry number is 
close to zero are most likely superpositions of an eigenvector of $D_{A}+A$ 
and an eigenvector of $D_{B}+B$ with eigenvalues lying close in the
unperturbed spectrum where $g=0$. Let the unperturbed spacing between the 
two eigenvalues be denoted $\Delta$ and consider the matrix which connect 
the two states when the symmetry-breaking perturbation is introduced:

\begin{equation}
  H_{c} = \ma{0}{c}{c}{\Delta} \ .
\end{equation}

\noindent
The distribution, $P_{C}$, of the matrix elements, $c$, is Gaussian with zero 
mean and variance $(g\sigma)^{2}/2$. In the nume\-ri\-cal simu\-lation shown 
in Fig.\ \ref{asymmetry_1:1.fig} the eigenvectors come from the ``middle'' of 
the 
eigenvalue spectrum where the level density is almost constant, and we thus 
assume that the level density for each diagonal block is equal to a constant
which we denote $R_{1}$. In this case the spacing, $\Delta$, comes from the 
distribution 

\begin{equation}
  P_{\Delta}(\Delta) = \frac{1}{\sqrt{2\pi} \Delta_{0} }
              \exp \left( \ -\frac{\Delta^2}{2\Delta_{0}^2} \ \right) \ ,
\end{equation}

\noindent
where the variance $\Delta_{0}^2=1/(2\pi R_{1}^{2})$. In the two-dimensional 
approximation the eigenvectors have the asymmetry numbers

\begin{equation}
  a = \pm \frac{1}{\sqrt{1+4 c^2 / \Delta ^2}} \ ,
\end{equation}

\noindent
and thus the distribution of the asymmetry numbers becomes

\begin{eqnarray}
  P_{A}(a) &=& \frac{1}{2}
           \int_{-\infty}^{\infty} \ud c \int_{-\infty}^{\infty} \ud \Delta
           \   P_{C}(c) P_{\Delta}(\Delta)  \\
      & & \hspace{-1cm} \times
 \left[\delta \left( a - \frac{1}{\sqrt{1+ 4c^2/\Delta^2}} \right)
            + \delta \left( a + \frac{1}{\sqrt{1+ 4c^2/\Delta^2}} \right)
              \right] \
          \nonumber \\
        &=& \frac{\sqrt{2} \sigma g}
    {\pi \Delta_{0}\sqrt{1-a^{2}}[1-a^{2}+(\sqrt{2}a\sigma g/\Delta_{0})^2]}
            \label{asym_int} \nonumber \ .
\end{eqnarray}

\noindent
When $a=0$ the expression reduces to 

\begin{equation}
\label{pa_dist}
  P_{A}(0) = \frac{2 \sigma R_{1}}{\sqrt{\pi}} g \ , 
\end{equation}

\noindent 
which is in perfect agreement with the numerical simulation shown in Fig.\ 
\ref{asymmetry_1:1.fig} for which $\sigma=4/300$ and $R_{1}=140$.

The majority of the eigenvectors, which have values of $|a| \approx 1$, can
be described using perturbation theory. The shift in the position of the peak
of the distribution away from $a=1$ is to first order 
proportional to $g^{2}$, since the average correction to a given state from a 
state from the other block is proportional to $g$.

\section{The Acoustic Experiment}
\label{acoustic.sec}

\subsection{Acoustic resonances in thin plates}
\label{acoustic_theory.sec}

In a homogeneous and
isotropic three-dimensional medium, sound waves obey the elastomechanical
wave equation for the vectorial displacement field ${\bf u}$:
\begin{equation}
\rho{\partial^2{\bf u} \over \partial t^2} =
(\lambda + \mu)\nabla(\nabla\cdot{\bf u}) + \mu\nabla^2{\bf u}\;,
\label{full_elast_eq}
\end{equation}
where $\lambda$ and $\mu$ are the Lam\'e coefficients, $\rho$ is the
density, and we have assumed no external forces. Equation 
(\ref{full_elast_eq}) allows for two types of wave motion:
longitudinal and transverse.  (In the literature the transverse modes have 
names like {\it shear} or {\it secondary}, and
the longitudinal modes are often called {\it pressure} or {\it primary}.) 
Longitudinal waves always travel faster than 
transverse waves. For aluminium, which we consider in this paper, the
difference is approximately a factor of 2. In the
bulk, the two types of waves propagate independently. However, upon
reflection at a boundary, mode conversion takes place: an incident
wave that is purely longitudinal or transverse will, in general, give rise
to {\it two} reflected waves, one longitudinal and one transverse. Moreover,
their angles of reflection will be different due to their different
velocities, as dictated by Snell's law.

We now briefly present some facts about elastic waves in thin infinite plates,
see, e.g., \cite{Graff:1975} and the recent studies in 
Ref. \cite{Hugues:2000,Bogomolny:1998}.
Three types of modes exist in an infinite isotropic plate, when considered at
frequencies below the first critical frequency, i.e., when one half of a
transverse wavelength is larger than the thickness of the plate.
The {\it flexural} modes have displacement mainly normal to the plane of
the plate, but they also have a small in-plane component. These modes are 
anti-symmetric with respect to reflection through the middle plane of the 
plate. (In the literature the 
flexural modes are sometimes called {\it bending}
modes.) The {\it in-plane} modes are symmetric with respect to reflection 
through the middle plane of the plate and consist of two mode types. The 
in-plane {\it transverse} modes 
have displacement exactly in the plane of the plate, and the in-plane 
{\it longitudinal} modes have displacement mainly in the
plane of the plate, but they also have a small out-of-plane component.

Now consider a finite plate. As mentioned above, the boundaries introduce 
mode conversion. For a finite plate there is thus the possibility of a 
coupling between the different mode classes. In Ref. \cite{Hugues:2000} it 
was concluded, first, that the flexural modes are un\-coupled from the 
in-plane modes and, second, that the in-plane longitudinal modes couple to 
the in-plane transverse modes. The densities of flexural modes and in-plane 
modes were 
calculated theoretically and found to be of the same order \cite{Hugues:2000}. 
These results explain the spectral fluctuation statistics measured in 
Ref. \cite{Hugues:2000} where resonances, i.e., both flexural and in-plane 
modes, of a quarter of a thin Sinai stadium plate were investigated. 
In Sec.\ \ref{separation_res.sec} we explain how to separate the flexural and 
in-plane modes experimentally using their measured widths. This technique 
allows us to measure the number 
of modes of the two types separately and to compare these numbers to the 
theoretical predictions of Ref. \cite{Hugues:2000}. It also enables us to 
study the spectral statistics and the width distributions for the two classes 
of modes independently and to find out if the flexural modes are in fact 
uncoupled from the other modes.

\subsection{Acoustic systems and experimental technique}
\label{acou_sys.sec}

\begin{figure}[h!]
  \epsfxsize=6.0cm
  \centerline{ { \epsffile{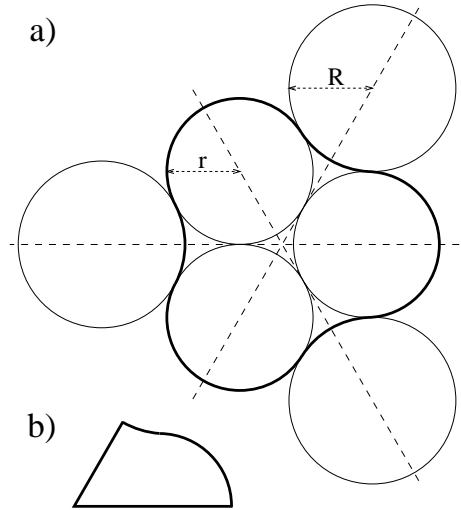} } }
  \vspace{0.2cm}
  \caption{Sketch of the three-leaf clover (not to scale). (a) The 
           construction of the three-leaf clover (thick solid line) with 
           mirror symmetries indicated by dashed lines. The geometry is 
           defined by the two radii, $r$  and $R$. (b) The plates used in 
           the experiments were one sixth of the three-leaf clover.}
  \label{clover.fig}
\end{figure}

For the experiments, we use two aluminium plates of different thickness 
cut in the shape of the three-leaf clover shown in Fig.\ \ref{clover.fig}.  
This billiard, which was first considered in Ref.\ \cite{Jarzynski:1993}, 
was chosen because it is known to be classically chaotic and, when $R \ge r$, 
it has no continuous families of periodic orbits \cite{Jarzyn_private}. Thus, 
we have chosen $r = 70$ mm and $R = 80$ mm. The area of the plates was 
$8250 \pm 100$ mm$^2$, and the circumference was $390 \pm 3$ mm. The plates 
were $1.5$ mm and $2$ mm thick.

The choices of $r$ and $R$ and the thickness
are important for the experiment in so far as they determine the re\-la\-tive 
densities of the two mode classes and also the total number of modes.
In our case, these parameters were chosen to give many modes for the
purpose of producing significant statistics while keeping the density of
in-plane modes approximately equal to the density of flexural modes in the
frequency range (300 kHz -- 600 kHz) where our transducers are most 
effective.
 
Aluminium was chosen for the plates because it is isotropic and very easy to
machine, while maintaining a high Q value; at 500 kHz the Q value measured 
in va\-cuum is around 10${}^4$. There are isotropic materials with much
higher Q values, such as fused quartz. However, fused quartz is more difficult 
to machine and thus not suitable for the symmetry-breaking 
experiment, where one must remove material from the plate many times in a 
controlled way.

The elastic constants for the two plates cannot be found in standard tables of
material properties, since they are not pure aluminium but a 
special alloy. However, the elastic constants for this alloy were determined 
by experiment in Ref. \cite{Bertelsen:1997}. We shall use the values 
from Ref. \cite{Bertelsen:1997} for Young's modulus $E =$70 $\pm$ 1 GPa and 
Poisson's ratio $\nu =$0.330 $\pm$ 0.005. The density is 2.698 g/cm${}^3$ 
\cite{Kaye:1986}. The corresponding bulk sound
velocities are 3123 m/s for transverse waves and 6200 m/s for longitudinal 
waves. 

The experimental setup is in many ways the same as that used for
previous experiments as reported in \cite{acoustic3:1996}. We use an HP 3589A
spectrum/network analyser to measure transmission
spectra of acoustic resonators via piezoelectric transducers. The plate
rests horizontally, supported by three
gramophone diamond styli. This ensures a very small contact area between
the plate and the rest of the world, thus making the vibrations of the plate
as close to free as possible. The diamond styli are glued to cylindrical
piezo ceramics that are polarised along the symmetry axis (z-axis). One such
combination functions as transmitter, the two others as receivers.

One may wonder if our experimental technique can really measure all modes. 
In particular, one could question if the in-plane modes, for which the 
displacement is mainly (or exactly) in the plane of the plate, are detected 
by our transducers. This question was answered in Ref. \cite{Hugues:2000}, 
where the same experimental technique is used. The authors find that all 
modes are detected. To understand
this, one can imagine what happens microscopically when strain is passed
from the plate to the piezoelectric component through the diamond stylus.
Obviously, there can be no slip between the tip of the stylus and the plate.
If there were indeed slip, there would also be friction. The diamond would 
then quickly drill a hole in the plate, and this is not observed in the 
experiments. In fact, after many days of oscillations at frequencies of 
several hundred kHz, the plate is completely intact. Since the base of the 
diamond stylus is fixed to the piezo electric component, the diamond stylus 
undergoes a wiggling motion which deforms the piezo electric component in a 
complicated way, including compression along the z-axis.

In both of the experiments the temperature was room temperature, i.e., it
was not kept constant but could fluctuate by a few degrees.
Obviously, the temperature is important in these measurements, since both
the size of the plate and the elastic constants depend on the temperature, and 
changes in these parameters affect the eigenfrequencies.  
However, for aluminium thermal expansion is the dominant effect, and to 
first order eigenfrequencies shift locally by the same amount. Since 
we are not interested in single eigenfrequencies but only in differences 
between them, this shift has no influence on our results.

\begin{figure}[here]
  \centerline{\epsfxsize=8.5cm
              \epsffile{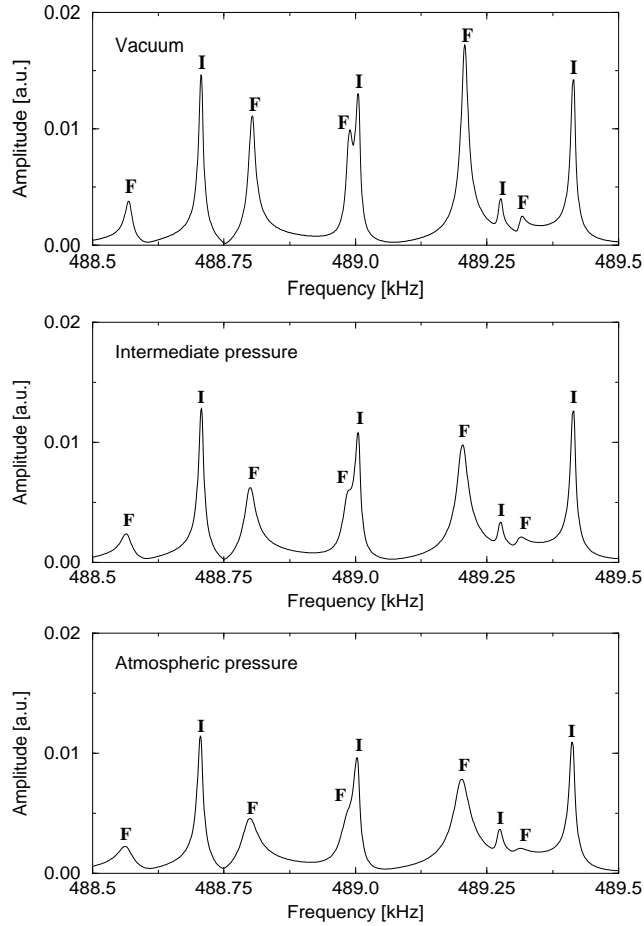}
             }
        \vspace{0.2cm}   
        \caption{A segment of the resonance spectrum for the 2 mm thick 
                 three-leaf clover 
                 at three different pressures: vacuum, 0.5 atm, and 1 atm. 
                 Most resonances are easily recognised to be either in-plane 
                 ({\bf I}) or flexural ({\bf F}).} 
        \label{spec_pressure.fig}

\end{figure}

The plate is placed in a vacuum chamber, which allows control of the
pressure of the air surrounding the plate. At pressures lower than $10^{-2}$ 
Torr air damping is insignificant compared to intrinsic losses and losses to 
the supports. Therefore, we shall refer to such low pressure
as ``vacuum''. When the pressure is increased, the flexural modes, that have 
large out-of-plane oscillations, are strongly affected, since the plate then 
functions like a loudspeaker generating sound waves in the air. As a result, 
the amplitudes of the flexural modes decrease with increasing pressure, and 
the widths of the resonance peaks increase. This is demonstrated in 
Fig.\ \ref{spec_pressure.fig}, which shows a section of the transmission 
spectrum measured for the three-leaf clover in vacuum, at a pressure of 
0.5 atm, and at atmospheric pressure. Note that one can label most of the 
modes into flexural and in-plane by eye.

\subsection{The separation experiment}
\label{separation_res.sec}

The first experiment was designed to separate the modes into flexural and 
in-plane types so that
the spectral statistics could be studied separately for each class. To get a
statistically significant result, many eigenfrequencies are needed, and it is
crucial to find all the levels so that the results are free from
missing level effects. For this reason we performed the following measurement
sequence. The acoustic transmission spectrum for the plate of thickness 2 mm 
was measured in the range 300 
kHz -- 540 kHz. The measurement was carried out first in vacuum, then at a
pressure of 0.5 atm, and finally at 1 atm, see Fig.\ \ref{spec_pressure.fig}. 
In each case, the measurement was
performed twice, using two independent receivers. This procedure gave 6
resonance spectra. Then, the system was subject to a perturbation, when a mass
of 14 mg, corresponding to 314 ppm, was removed from one face of the plate 
using a piece of fine sand paper. After this, the above procedure was repeated,
giving another 6 resonance spectra. Then, in the same way, another 
perturbation was made, this time removing 43 mg of material, corresponding to
965 ppm. Again, the measurement sequence was carried out, giving a total of
18 resonance spectra. The perturbations done to the system are small enough
that it is possible to follow every resonance peak through all 18 spectra, but
large enough that near-degeneracies in one set of spectra are destroyed by
the perturbations, giving well-resolved peaks in the next set of spectra. This
technique allows us to find all resonances. There are no missing levels.

\begin{figure}[here]
 \centerline{\epsfxsize=8.5cm
              \epsffile{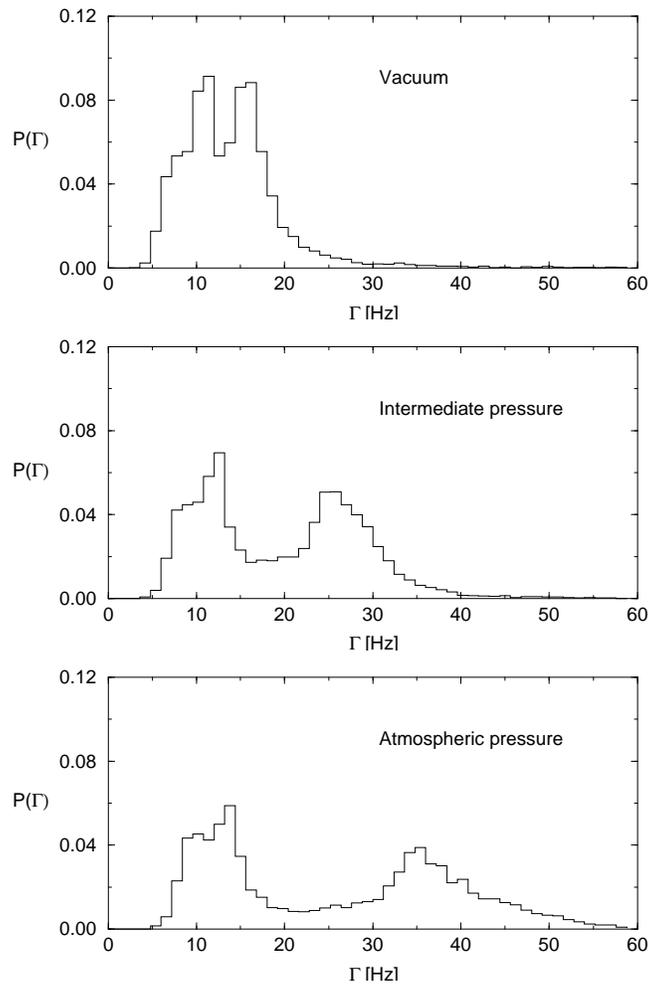}
             }
  \vspace{0.2cm}   
  \caption{The distribution of widths for the acoustic resonances for 
           the three-leaf clover plate of thickness 2 mm, at the same
           values of the air-pressure as in Fig.\ \ref{spec_pressure.fig}. 
           One class of modes (in-plane) have widths which are almost 
           independent of the pressure and the widths for the other class 
           of modes (flexural) increase with increasing pressure.}
        \label{width_pressure.fig}
\end{figure}

We would like to establish a simple and reliable criterion that permits 
us to separate the spectrum into flexural and in-plane modes.  To this 
end, each resonance peak is fitted using the so-called ``skew Lorentzian'' 
approach \cite{Alt:1993}.  This fit yields a number of parameters of which 
only the resonance frequency and the width, $\Gamma$, are of interest. In 
Fig.\ \ref{width_pressure.fig} 
we show the distribution of widths obtained from this fitting procedure 
for increasing values of the air pressure.  It is evident that the widths 
of one group of modes increases with increasing pressure while the widths 
of the remaining modes is largely unaffected.  We interpret these groups as 
flexural and in-plane modes, respectively.  However, even at atmospheric 
pressure, it is not possible to separate the modes on the basis of 
resonance width alone.

\begin{figure}[h!]
  \epsfxsize=8.5cm
  \centerline{\epsffile{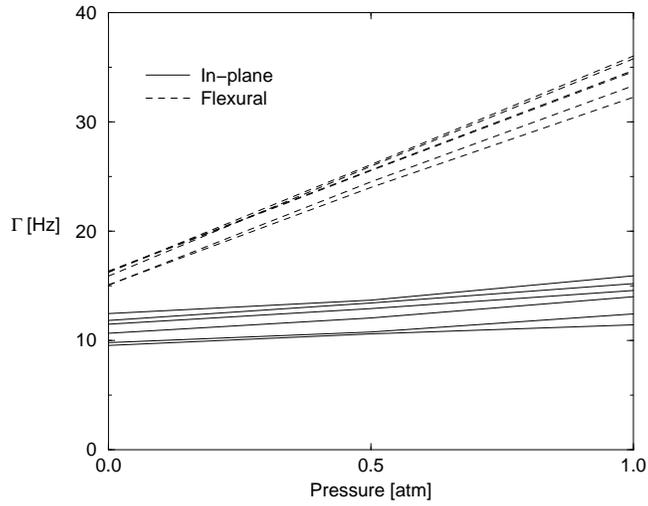} } 
  \vspace{0.2cm}
  \caption{The width of a flexural mode and an in-plane mode for the
           three-leaf clover as function of the pressure of the air in the 
           vacuum chamber. Six curves appear for each resonance peak, because
           six different measurements were made for each of the pressure 
           values: vacuum, 0.5 atm, and 1 atm.}
  \label{width_of_p.fig}
\end{figure}

Since the width distribution does not allow us to separate the flexural
modes from the in-plane modes with certainty, we must find a more reliable
criterion. Therefore, we consider the individual resonance widths as
function of pressure, see Fig.\ \ref{width_of_p.fig}.

\begin{figure}[here]
        \centerline{             
                    \epsfxsize=8.5cm
                    \epsffile{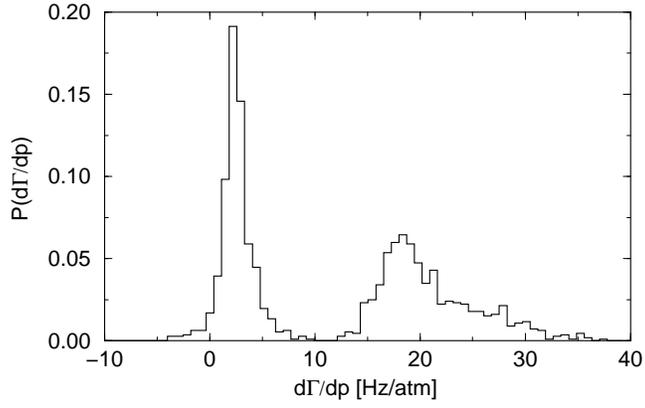} 
                   }
        \vspace{0.2cm}   
        \caption{The distribution of slopes $\ud\Gamma/\ud p$ has two 
                 well-separated peaks, which makes it possible to separate
                 the flexural and in-plane modes. We choose a ``separation'' 
                 slope of 11 Hz/atm. A few inaccurate fits gives rise to the 
                 small number of negative slopes.}
        \label{dw_dp_dstr.fig}
\end{figure}

The curves for the two resonances in Fig.\ \ref{width_of_p.fig} are typical 
for the measured 
modes and show that the curves are well approximated by straight lines.
Consequently, it makes sense to label them by the slope of the best 
straight line fit. We then consider the distribution of these slopes, see 
Fig.\ \ref{dw_dp_dstr.fig}. The distribution has two well-separated peaks. 
Large slopes correspond to flexural modes; small slopes correspond to in-plane 
modes. Based on this information, we choose the ``separation'' slope to be 
11 Hz/atm.

In the range 300 kHz to 540 kHz we find 1537 levels for the 2 mm plate, of 
which 781 are flexural 
and 756 are in-plane, judging from the sepa\-ration criterion discussed above.
Reference \cite{Hugues:2000} presents an expansion of the exact dispersion
relations for an infinite isotropic plate and also gives the corresponding
expansion for the number of modes, i.e., the {\it staircase function}, for a
finite, thin plate. Using this theoretical expansion, we expect $782$ flexural 
modes and $753$ in-plane modes. This is in perfect agreement with the measured 
numbers, given the uncertainty in the elastic constants of the aluminium alloy 
and in the dimensions of the plate.

\begin{figure}[here]
\centerline{\epsfxsize=8.5cm
            \epsfysize=12.5cm
            \epsffile{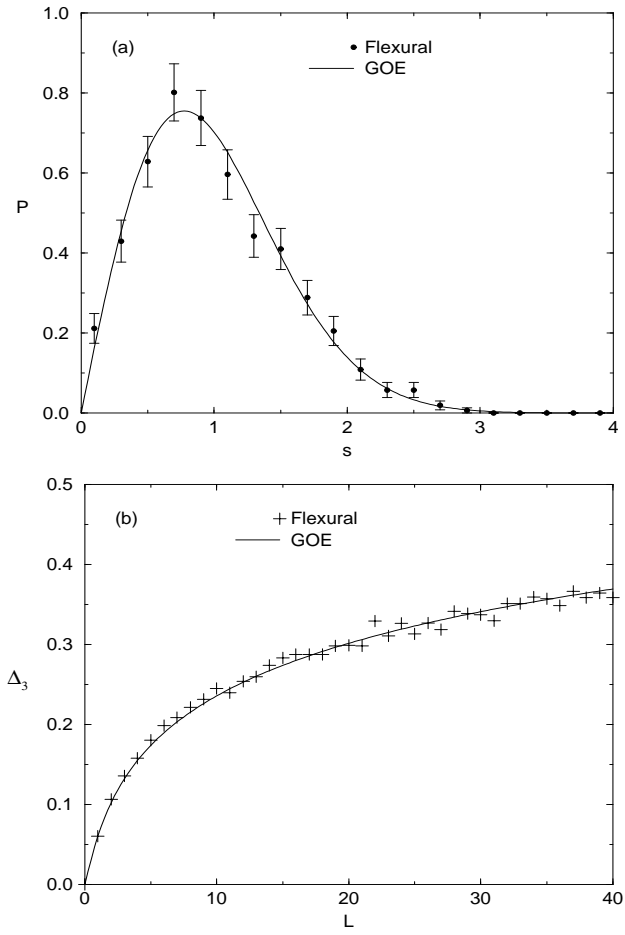}
           } 
           \vspace{0.2cm}
  \caption{The level spacing distribution (a) and the $\Delta_3$-statistic 
           (b) for the flexural modes compared with the GOE statistics. Both
           the level spacing distribution and the $\Delta_3$-statistic
           agree very well with the GOE results.}
  \label{fl_delta3.fig}
\end{figure}

\begin{figure}[here]
\centerline{\epsfxsize=8.5cm
            \epsfysize=12.5cm
            \epsffile{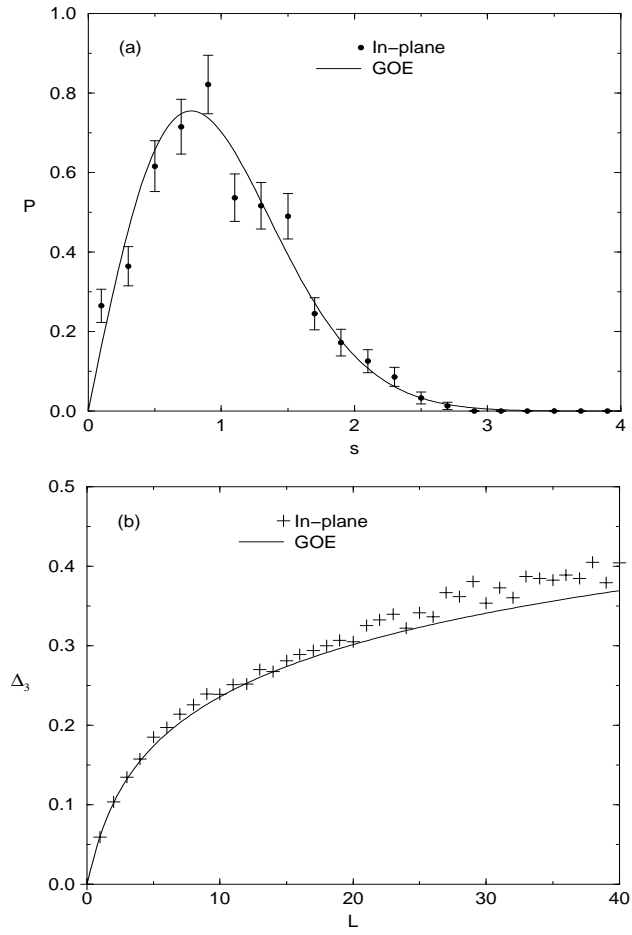}
           }
           \vspace{0.2cm}
  \caption{The level spacing distribution (a) and the $\Delta_3$-statistic 
           (b) for the in-plane modes compared with the GOE statistics. The 
           $\Delta_3$-statistic shows that the spectrum of the in-plane 
           modes is slightly less rigid than the GOE.}
  \label{ip_delta3.fig}
\end{figure}

Since we can identify the character of individual modes, it is possible to 
consider the level spacing distribution and the $\Delta_3$-statistic 
separately for each of the two classes. Figures \ref{fl_delta3.fig} and 
\ref{ip_delta3.fig} show the level spacing distribution and the 
$\Delta_3$-statistic for each of the two mode classes compared with  
the GOE statistics. We find that both the level spacing distribution and the 
$\Delta_3$-statistic for the flexural modes agree with the GOE statistics. 
This result confirms numerical calculations by Bogomolny and Hugues 
showing that the flexural modes of a chaotic billiard have GOE fluctuation 
statistics, see Ref. \cite{Bogomolny:1998}. The $\Delta_3$-statistic for the 
in-plane modes lies above the GOE curve. This is a bit surprising, because
mode conversion is expected to be a strong effect, see, 
e.g. Ref.\ \cite{Couchman:1992}, which should guarantee that 
all in-plane modes are strongly coupled and obey GOE statistics. 
We note that the deviation from the GOE curve seen in the $\Delta_3$-statistic
does not appear in the spacing distribution; the spacing 
distribution for the in-plane modes looks much like the spacing distribution 
for the GOE. The same feature is seen for the random matrix 
model for systems with an approximate symmetry, see the results for $g=0.2$ on 
Figs. \ref{levelspacing.fig}(c) and \ref{delta3.fig}. If we think of 
mode conversion as a mechanism that breaks the longitudinal-transverse 
``symmetry'' for in-plane modes, our results could indicate that this symmetry 
is not completely broken. 

An issue to consider in this context is the value of the wavelength, $\lambda$,
compared to the size, $l$, of the system. The ratio $l / \lambda$ is a measure 
of how ``semiclassical'' our system is. Roughly, $l=100$ mm. Random 
matrix results are only expected to apply when $l / \lambda  \gg 1$. For 
flexural 
modes, the typical wavelength is 5 mm, so $l / \lambda = 20$. For travelling
in-plane waves, the typical transverse wavelength is 7 mm and the typical
longitudinal wavelength is 13 mm. Roughly, this leads to $l / \lambda = 10$.
Thus, in our experiments we have the two length scales separated by at least 
an order of magnitude. Nevertheless, the factor of 2 between $l / \lambda$ for
flexural and in-plane modes shows that the flexural modes are more
``semiclassical'' than the in-plane modes, which is another possible 
explanation for the slight difference observed in the fluctuation properties. 

We emphasise that the main results of this section are, first, that the 
flexural and the in-plane modes can be separated and, second, that each of 
the two mode classes behave as one class of strongly-coupled modes. The fact 
that the $\Delta_3$-statistic lies slightly above the GOE curve for the 
in-plane 
modes is a small correction to this picture. In the following section, we 
regard the in-plane modes as one class of strongly-coupled modes.

\subsection{The symmetry-breaking experiment}
\label{res_symm_break.sec}

\begin{figure}[h!]
  \epsfxsize=7.00cm
  \centerline{ \epsffile{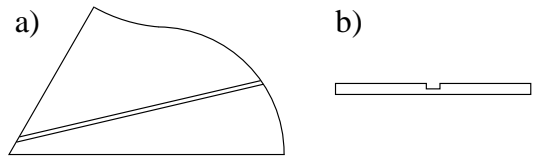} }
  \vspace{0.2cm}
  \caption{(a) A sixth of the three-leaf clover with a 2 mm wide cut
           indicated (not to scale). (b) Sketch of the plate profile
           showing how the cut breaks the mirror-symmetry through the middle
           plane of the plate.}
  \label{clover_slit.fig}
\end{figure}

The second experiment was designed for a detailed study of the transition
from two independent mode classes to one mode class. The transition takes 
place as the mirror-symmetry through the middle plane of the plate is broken.
For this experiment, we used the three-leaf clover plate of thickness 1.5 mm
and gradually cut a slit on one side of the plate, as shown in 
Fig.\ \ref{clover_slit.fig}.

For the cutting of the slit in the plate we used a computer-controlled milling 
machine and chose steps in the thickness of 1/40 mm. In our case, 
this amounted to about 18 mg of material for each increment of the depth of the
slit. The mass of the intact plate was 32.8870 g. First, the frequency spectrum
was measured for the intact plate in vacuum and at atmospheric pressure.
The procedure of cutting and measuring the frequency spectrum at
atmospheric pressure was then repeated 9 times. In all measurements the 
frequency range was 456 kHz -- 533 kHz and only one receiver was used. The 
justification for using just one receiver for this experiment is as follows:
Removal of material from the plate corresponds to a small perturbation.
One can therefore easily follow each resonance peak through the entire 
scenario, and although a resonance peak can sometimes disappear in one 
spectrum because the receiver is accidentally placed on a nodal line, it 
always reappears in subsequent spectra. Thus, the results of this
symmetry-breaking experiment are protected against missing level effects. 

As in the previous experiment, the resonance peaks are fitted and we calculate 
the distribution of widths, focusing first on the intact plate. In the plot 
for atmospheric pressure, the modes are separated into two classes: those 
that have widths smaller than 22 Hz and those that have widths larger than 
22 Hz, see Fig.\ \ref{sym_break_width.fig}. This sets the criterion for 
sepa\-ration of the flexural modes from 
the in-plane modes. We note that for the 1.5 mm plate it is possible to 
perform the separation purely on the basis of the widths measured at 
atmospheric pressure. This was not the case for the 2 mm plate. 

In general, we expect that the widths of the flexural modes at some value
of the pressure will depend on many parameters. Among these, the thickness 
of the plate and the typical wavelength play important roles. However,
comparing our two experiments, all of the parameters are the same except for 
the thickness. The average width for the in-plane modes is about the same in 
the two cases. At a pressure of 1 atm, the mean width for the flexural 
modes for the 2 mm plate is around 35 Hz and for the 1.5 mm plate the 
mean width is 42 Hz. This indicates that damping from the air is larger for 
thinner plates.

We consider first the plate before any material has been removed and find 600 
levels in the frequency range 456 kHz - 533 kHz. 
According to our sepa\-ration rule, this time based solely on the width 
distribution measured at atmospheric pressure, 310 modes are flexural and 
290 are in-plane. Using again the expansion for the number of modes given in 
Ref.\ \cite{Hugues:2000}, there should be 311 flexural modes and 285 in-plane
modes, in perfect agreement with our results. As in Sec. 
\ref{separation_res.sec} we have obtained the level spacing di\-stribution 
and the $\Delta_3$-statistic for the two mode classes separately. We 
find the same spectral statistics for the 1.5 mm plate as for the 2 mm plate.

\begin{figure}[here]
           \centerline{
                       \epsfxsize=7.0cm
                       \epsffile{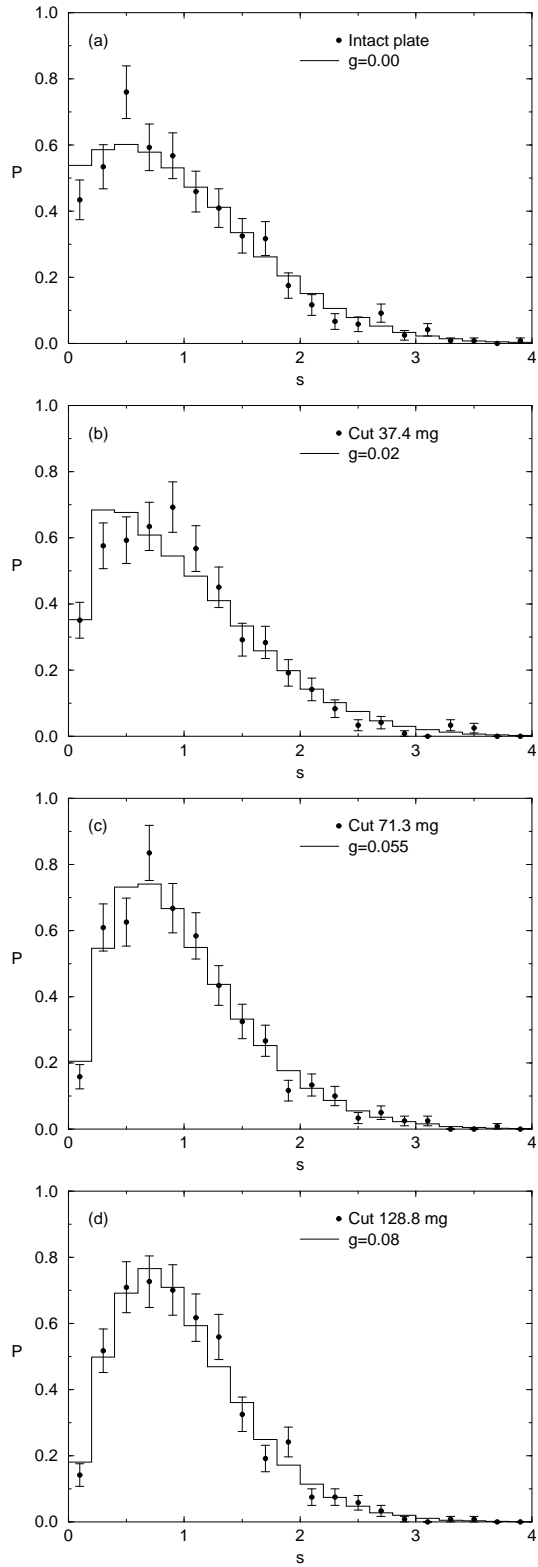}
                      }
  \caption{The level spacing distribution for all the modes compared to the 
           result of the random matrix model. 
           (a) The plate is intact. 
           (b) A total of 37.4 mg has been removed.
           (c) A total of 71.3 mg has been removed.
           (d) A total of 128.8 mg has been removed.}
  \label{sym_break_ls.fig}
\end{figure}

\begin{figure}[here]
           \centerline{
                       \epsfxsize=7.0cm
                       \epsffile{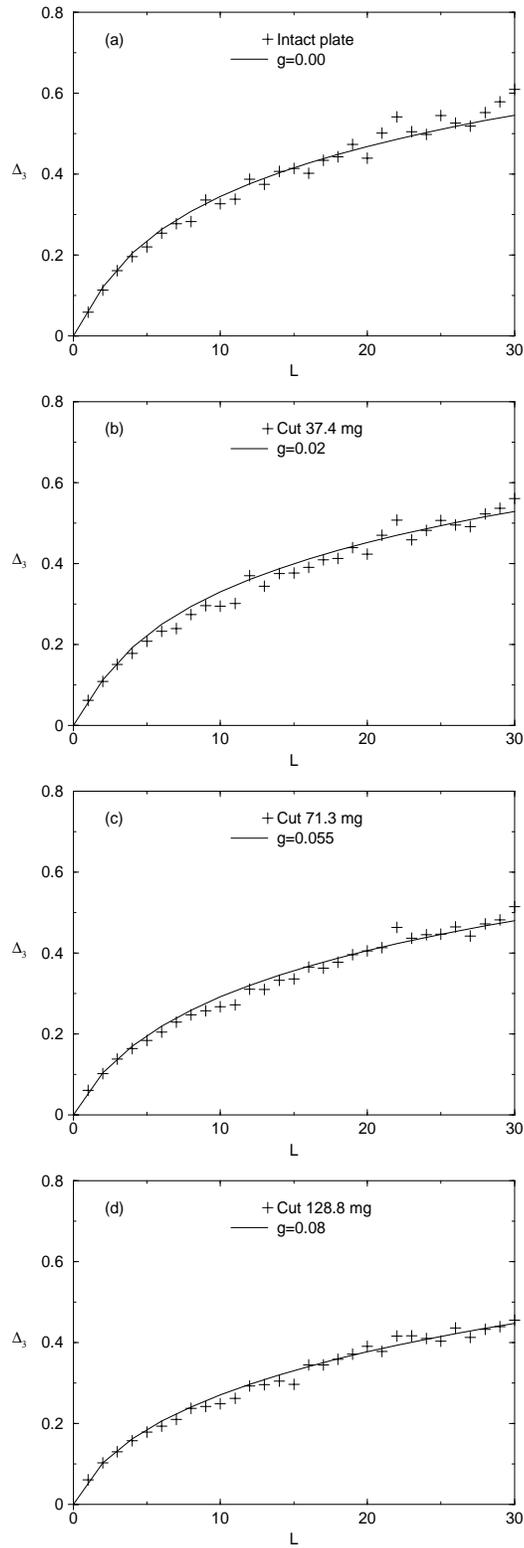}
                      }   
  \caption{The $\Delta_{3}$-statistic for all the modes is fitted by 
           minimising the sum of the squared deviations between the 
           data and the random matrix result. 
           (a) The plate is intact. 
           (b) A total of 37.4 mg has been removed.
           (c) A total of 71.3 mg has been removed. 
           (d) A total of 128.8 mg has been removed.}
  \label{sym_break_delta3.fig} 
\end{figure}

\begin{figure}[here]
           \centerline{
                       \epsfxsize=7.0cm
                       \epsffile{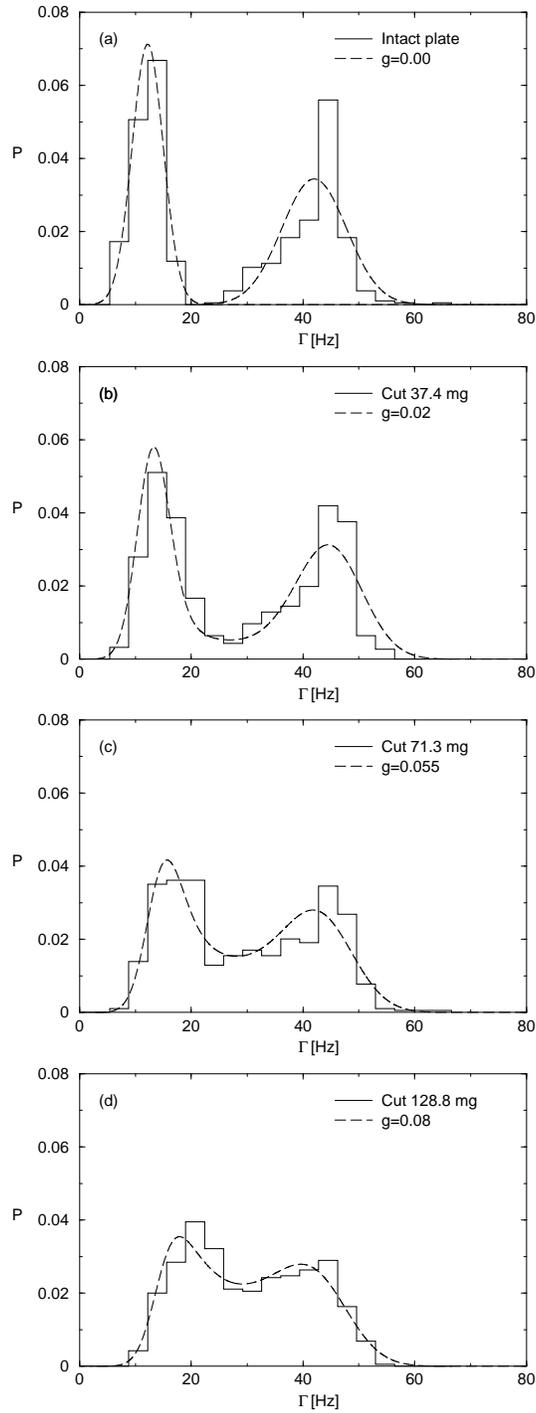}
                      } 
  \caption{The experimental width distribution for the 
           resonances compared to the distribution $P(\GA )$ obtained 
           numerically using Eq. (\ref{width.eqn}). The three plots
           represent data for the same slit-depths and values of $g$ as in 
           Figs.\ \ref{sym_break_ls.fig} and \ref{sym_break_delta3.fig}.
           (a) The plate is intact. 
           (b) A total of 37.4 mg has been removed.
           (c) A total of 71.3 mg has been removed.
           (d) A total of 128.8 mg has been removed.} 
  \label{sym_break_width.fig}
\end{figure}

Figures \ref{sym_break_ls.fig} and \ref{sym_break_delta3.fig} show the level 
spacing di\-stributions and the $\Delta_3$-statistics for all the modes for 
increasing depth of the symmetry-breaking slit. The experimental data are
fitted with results for the random matrix model of Sec. \ref{theory.sec}.
We have used $N_1 = N_2 = 150$ and $\sigma^2=64/N^2$. Table 
\ref{symmetry.table} summarises the results for the theoretical fits to the
spectral statistics for 
the symmetry breaking experiment. The spectral statistics are well described 
by the model, and the best fits to the level spacing distribution and to the
$\Delta_3$-statistic yield consistent values for the coupling strength $g$.

In Fig.\ \ref{sym_break_width.fig} the measured width distributions are 
compared with the distributions calculated nume\-rically using Eq. 
(\ref{width.eqn}). To model the width distribution $P(\GA )$ for all modes, 
we use the asymmetry distribution for the eigenvectors and assume that the 
in-plane modes and the flexural modes have Gaussian width distributions 
$P_{I}$ and $P_{F}$. For each of the different cases we fix the value of the 
coupling strength, $g$, to the value obtained from the spectral statistics.   
For the intact plate, see Fig. \ref{sym_break_width.fig}(a), we fitted the 
width distribution by minimising $\chi^2$, and found the mean va\-lues 
$\GA_{I}^{0}$ = 12.2 Hz and $\GA_{F}^{0}$ = 42.0 Hz, and the standard 
deviations $\sigma_{I}$ = 2.8 Hz and $\sigma_{F}$ = 5.8 Hz. The mean values 
for the fit agree with the measured average width of 27.1 Hz, see 
Tab. \ref{symmetry.table}. The average width depends on the slit depth
as shown in Tab. \ref{symmetry.table}, and increases, e.g., 
by 2.1 Hz when the cut increases from 0 mg to 37.4 mg. To take effects like
this into account we have fitted the width
distributions by varying the four parameters $\GA_{I}^{0}$, 
$\GA_{F}^{0}$, $\sigma_{I}$, and $\sigma_{F}$. The only parameter which changed
significantly from case to case was the average width of the flexural 
resonances, $\GA_{F}^{0}$. This seems reasonable since we expect that the
modes with large out-of-plane components are damped most by surface 
perturbations like the cut. For the width distributions shown in Fig. 15(b), 
(c), and (d) we therefore held $\GA_{I}^{0}$, $\sigma_{I}$, and 
$\sigma_{F}$ fixed, whereas $\GA_{F}^{0}$ was varied so that the 
average width equalled the measured average shown in Tab. \ref{symmetry.table}.
The overall features of the width distribution as function of slit depth
are described by the random matrix model. As the slit depth increases, the
strength of the width distribution between the two peaks increases while 
the strength of the peaks decreases. Notice that the value of $P(\GA )$ around 
$\Gamma=27.5$ Hz increases linearly with $g$ in agreement with Eq. (8). 
 
\section{Discussion and Conclusions}
\label{discuss_concl.sec}

We have presented experimental results for acoustic resonances in two thin
aluminium plates of three-leaf clover shape. For both plates
we found that the measured number of flexural and in-plane resonances were
in very good agreement with the theoretical Weyl formula. The two classes of
modes were separated using their width or the dependence of the width on the
pressure of the air surrounding the plate. The spectral statistics for the
flexural modes were in perfect agreement with the GOE result in both cases 
whereas the spectra of the in-plane modes seemed to be slightly less rigid 
than the GOE.

The random matrix model of systems with an 
approximate symmetry modelled the experimental data on the spectral
statistics and wave function information from the mixing experiment well. Both 
the level spacing distribution, the $\Delta_3$-statistic, and the 
distribution of widths were fitted consistently by the numerical random 
matrix results. The qualitative changes in the width distribution as the 
depth of the cut was increased could thus be ascribed to the complex mixed 
nature of the acoustic wave functions.

The successful description of the statistics of the frequency spectrum and 
the widths of the thin acoustic plates may be extended to include other
features. The presence of both a kinetic energy term and an interaction 
term in the random matrix model is natural not only in the modelling of the
mixing process but also to describe the Thouless energy of acoustic
resonators due to the localisation of wave functions. In this way the model 
represents an extension of the simplest random matrix models, like the 
GOE, to include several important features present in real physical systems.   

\section*{Acknowledgements}
\label{acknowledgements.sec}

KS acknowledges financial support from the Danish National Research Council.

\newpage

\begin{table}[h]
\begin{center}  
  \caption{The first column shows the size of the cut and the fourth column 
           the average width.
           The second and third column show estimates of $g$ obtained by
           fitting the level spacing distribution and the 
           $\Delta_{3}$-statistic, respectively.}
  \vspace{0.2cm}
  \begin{tabular}{rllc}
  Cut [mg] & $g$ from ls   
           & $g$ from $\Delta_{3}$ 
           & Mean width [Hz] \\
    \hline
    0.0 & 0.01  & 0.00  & 27.1 \\
   10.3 & 0.02  & 0.00  & 27.9 \\
   23.1 & 0.02  & 0.005 & 28.6 \\ 
   37.4 & 0.02  & 0.02  & 29.2 \\
   55.7 & 0.05  & 0.04  & 29.1 \\
   71.3 & 0.055 & 0.055 & 29.7 \\
   92.5 & 0.06  & 0.05  & 30.0 \\
  108.5 & 0.06  & 0.06  & 30.8 \\
  128.8 & 0.09  & 0.08  & 30.3 \\
  146.1 & 0.06  & 0.07  & 29.9 \\ 
  \end{tabular}
  \label{symmetry.table}
\end{center}
\end{table}

\newpage 

\end{document}